\documentclass[twoside,journal]{IEEEtran}
\usepackage{amsmath,amsfonts}
\usepackage{algorithmic}
\usepackage{algorithm}
\usepackage{array}
\usepackage[caption=false,font=normalsize,labelfont=sf,textfont=sf]{subfig}
\usepackage{textcomp}
\usepackage{stfloats}
\usepackage{url}
\usepackage{verbatim}
\usepackage{graphicx}
\usepackage{cite}
\begin{document}
\title{Observation of Periodic Optical Spectra and Soliton Molecules in a Novel Passively Mode-Locked Fiber Laser}

\author{Xiang Zhang, Haobin Zheng, Kangrui Chang, Yong Shen, Yongzhuang Zhou, Qiao Lu, and Hongxin Zou 
\thanks{Manuscript received January 21, 2024; revised March 06, 2024. This work was supported in part by National Natural Science Foundation of China under Grants 62105368, 62275268 and 62375284, in part by Natural Science Foundation of Jiangsu Province under Grant BK20230418, and in part by The Science and Technology Innovation Program of Hunan Province under Grant 2023RC3010. {\it (Xiang Zhang and Haobin Zheng contributed equally to this work.)} {\it (Corresponding author: Qiao Lu and Hongxin Zou.)}} 
\thanks{Xiang Zhang, Haobin Zheng, Kangrui Chang, Yong Shen, Yongzhuang Zhou and Hongxin Zou are with Institute for Quantum Science and Technology, College of Science, National University of Defense Technology, Changsha 410073, China, and also with Hunan Key Laboratory of Mechanism and Technology of Quantum Information, Changsha 410073, China (e-mail: zhangxiang\_zx@nudt.edu.cn; zhenghaobin@nudt.edu.cn; changkangrui@foxmail.com; shenyong08@nudt.edu.cn; y.zhou@nudt.edu.cn; hxzou@nudt.edu.cn).} 
\thanks{Qiao Lu is with School of Automation, Nanjing University of Information Science and Technology, Nanjing 210044, China (e-mail: 003706@nuist.edu.cn).}}

\markboth{Journal of Lightwave Technology, ~Vol.~XX, No.~X, January~2024}%
{Zhang \MakeLowercase{\textit{et al.}}: Observation of Periodic Optical Spectra and Soliton Molecules}


\maketitle

\begin{abstract}
Due to the necessity of making a series of random adjustments after mode-locking in most experiments for preparing soliton molecules, the repeatability of the preparations remains a challenge. Here, we introduce a novel all-polarization-maintaining erbium-doped fiber laser that utilizes a nonlinear amplifying loop mirror for mode-locking and features a linear shape. This laser can stably output soliton molecules without any additional adjustment once the mode-locking self-starts. Moreover, it can achieve the transition from soliton molecule state to soliton state, and then to multi-pulse state by reducing the pumping power. The unconventional method of generating multi-pulses, combined with a wide pumping power range of 200--640 mW for maintaining mode-locking, allowed us to observe periodic optical spectra with two complete cycles for the first time. Based on the experimental facts, we develop a multistability model to explain this phenomenon. With its ability to switch between three stable states, this flexible laser can serve as a versatile toolbox for studying soliton dynamics.
\end{abstract}


\section{Introduction}
\IEEEPARstart{T}{hanks} to their outstanding performance, ultrashort pulses have been widely used in many fields, such as materials processing\cite{sugioka_will_2021}, cell imaging and gene editing\cite{denk_two-photon_1990,song_femtosecond_2023,xue_ultrasensitive_2019,zheng_highly_2022,chen_crisprcas12a-empowered_2022}, precision ranging\cite{caldwell_time-programmable_2022}, trace gas sensing\cite{barik_selectivity_2022}, quantum information and quantum computing\cite{menicucci_one-way_2008,pysher_parallel_2011}, frequency metrology\cite{inaba_long-term_2006}, and timing synchronization\cite{kim_drift-free_2008,shen_free-space_2022}. Because of its advantages such as portability, compact structure, ease of assembly, low cost, and resistance to environmental interference, the all-polarization-maintaining (PM) fiber mode-locked laser is often employed for generating ultrashort pulses, and its mode-locking is generally based on either real saturable absorbers such as semiconductor saturable absorber mirror (SESAM)\cite{keller_coupled-cavity_1990} or two types of artificial saturable absorbers, namely, nonlinear polarization rotation (NPR)\cite{noske_subpicosecond_1992,tamura_self-starting_1992} and nonlinear amplifying loop mirror (NALM)\cite{fermann_nonlinear_1990}. For the mode-locked fiber lasers, the linear cavity or linear shape has certain advantages in achieving a high repetition rate due to the avoidance of losses caused by fiber bending. By using a linear cavity, the all-PM erbium-doped fiber lasers based on SESAM and NPR can achieve the highest repetition rates of about 1 GHz and 115 MHz, respectively\cite{song_all-polarization-maintaining_2019,song_theoretical_2023,liu_115-mhz_2021}. These repetition rates are higher than those of the same kind lasers with the fiber loop. In contrast, the inclusion of an optical fiber loop is necessary for all-PM erbium-doped fiber lasers utilizing NALM due to their mode-locking mechanism, limiting their highest repetition rate to approximately 250 MHz\cite{hansel_all_2017}. If a technical method could be proposed to stretch the fiber loop of NALM into a linear shape, it may be possible to further increase the highest repetition rate.

In soliton dynamics research, it is common to prepare stable soliton molecules in non-PM fiber lasers. After mode-locking, some optimizations of the cavity parameters, such as adjusting polarization controllers, are necessary\cite{cui_dichromatic_2023,zhou_buildup_2020,zou_synchronization_2022}. However, once the lasers are restarted, the soliton molecules will disappear, and the processes of preparing soliton molecules need to be repeated. Due to changes in the cavity parameters, the new soliton molecules will likely differ from the old ones. Therefore, the repeatability of these experiments remains a challenge, and it is not helpful to obtain consistent conclusions.

The mode-locking of all-PM NALM-based fiber laser typically self-starts at high pumping power. A stable single pulse state can be obtained by gradually decreasing the power. When the pumping power exceeds a certain threshold, the single-pulse operation becomes unstable and a multi-pulse or noise-like pulse regime can be realized. This phenomenon in passive mode-locked lasers is known as pulse splitting\cite{komarov_pulse_2000}. However, the pulses in this regime generally lack long-term stability, and the spectral changes show a degree of randomness with variations in pumping power. Moreover, adjusting the pumping power and intracavity dispersion can change the number and time-domain distribution of multi-pulses\cite{laszczych_mode-locking_2022,laszczych_three_2022}. As the original multi-pulse configuration is disrupted, it makes it difficult to discover the underlying regularities. If stable multi-pulses can be generated at lower or near-soliton pumping powers with reduced nonlinear effects and spectral disturbances, it will facilitate the observation of new experimental phenomena in multi-pulse dynamics research.

In NPR-based ring-cavity fiber lasers, the spectra of dissipative solitons and noise-like pulses generally show regular changes as the pumping power monotonically increases or decreases\cite{lin_generation_2019,cheng_multi-shuttle_2020}. At high pumping power levels, a transition can occur between dissipative solitons and noise-like pulses, and accompanied by the reappearance of the vanished spectral profiles\cite{cheng_multi-shuttle_2020}. The self-similarity of dissipative solitons makes it difficult to determine whether the phenomenon itself is periodic. In soliton lasers with large net negative dispersion, the spectrum lacks self-similarity, but so far, no periodic spectrum has been observed.

This paper presents an all-PM NALM-based mode-locked fiber laser with a linear shape (not a linear cavity) consisting of two polarization beam combiner collimators (PBCCs), two fibers, and a free-space optical path. Once started at the appropriate pumping power, this laser can output soliton molecules without additional adjustments. Subsequently, reducing the pumping power can switch to soliton and multi-pulse states. The optical spectra of multi-pulses exhibit minimal distortion because of the low pumping power and show periodic changes as the pumping power is monotonically reduced. Based on existing experimental evidences, we propose a multistability model to interpret this phenomenon.

\section{Experiment setup and working principle}
\begin{figure}[!t]
\centering
\includegraphics[width=0.45\textwidth]{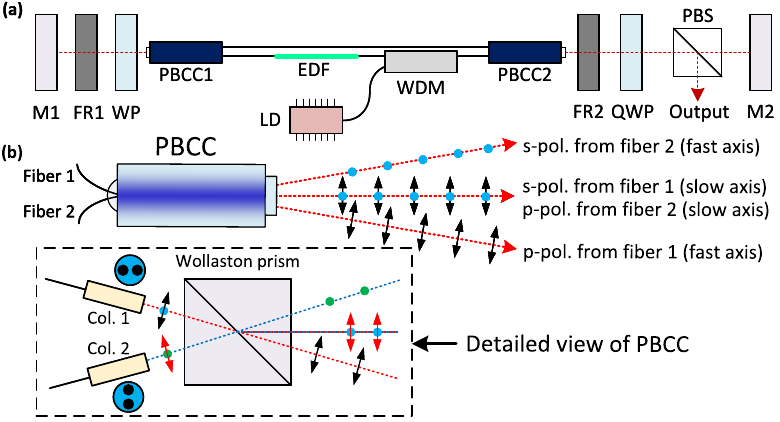}
\caption{(a) Experimental setup of the mode-locked fiber laser with a linear shape. M1 and M2, Mirror; WP, wave plate; EDF, erbium-doped fiber; WDM, wavelength division multiplexer; FR1 and FR2, Faraday rotator; LD, 976 nm laser diode; PBCC1 and PBCC2, polarization beam combiner collimator; PBS, polarizing beamsplitter. (b) The schematic of the polarization beam combiner collimator. Col. 1 and Col. 2, collimators; s-pol. and p-pol., s-polarized and p-polarized lights.}
\label{fig_1}
\end{figure}

\noindent The schematic of the linear-shape all-PM erbium-doped fiber laser mode-locked by a NALM is shown in Fig.~\ref{fig_1}(a). There are two 2.18-meter fiber segments, one is a passive PM fiber (PM1550) and the other is composed of passive PM fiber (PM1550) and 0.55-meter erbium-doped fiber (Er80-4/125-HD-PM, Liekki). The group velocity dispersion (GVD) of passive PM fiber is -22.8 ps$^2$/km at 1550 nm, and that of erbium-doped fiber is 28.04 ps$^2$/km at 1550 nm, and the net dispersion in the cavity is -71446 fs$^2$. As shown in Fig.~\ref{fig_1}(b), the PBCC (SR11141D, AFR) is constructed of a Wollaston prism and two gradient index (GRIN) lens collimators, and the polarization directions of the lights coupled with PBCCs (PBCC1 and PBCC2) and mirrors (M1 and M2) are along the slow axis of the two fibers. The Faraday rotator (FR1) and M1 on the left are used to couple the light output from one fiber into another. This cancels out the linear phase shifts of the lights in the two fibers and only enables the nonlinear phase shifts (NPSs) of the lights polarized along the slow axis of the fibers to affect the round-trip transmission of NALM when interfering on the polarizing beamsplitter (PBS). The Faraday rotator (FR2) and the quarter-wave plate (QWP) on the right form the phase shifter, which enables the mode-locking to self-start. In the experiment, because the extinction ratio of the PM fiber is less than 50 dB and the intracavity light is not purely linearly polarized, the PBCCs also output poor-quality weak lights polarized along the fast axis of the fiber, including continuous light generated by amplified spontaneous emission (ASE) and pulse light. Therefore, these output lights are not discussed in this paper.

\begin{figure}[!t]
\centering
\includegraphics[width=0.43\textwidth]{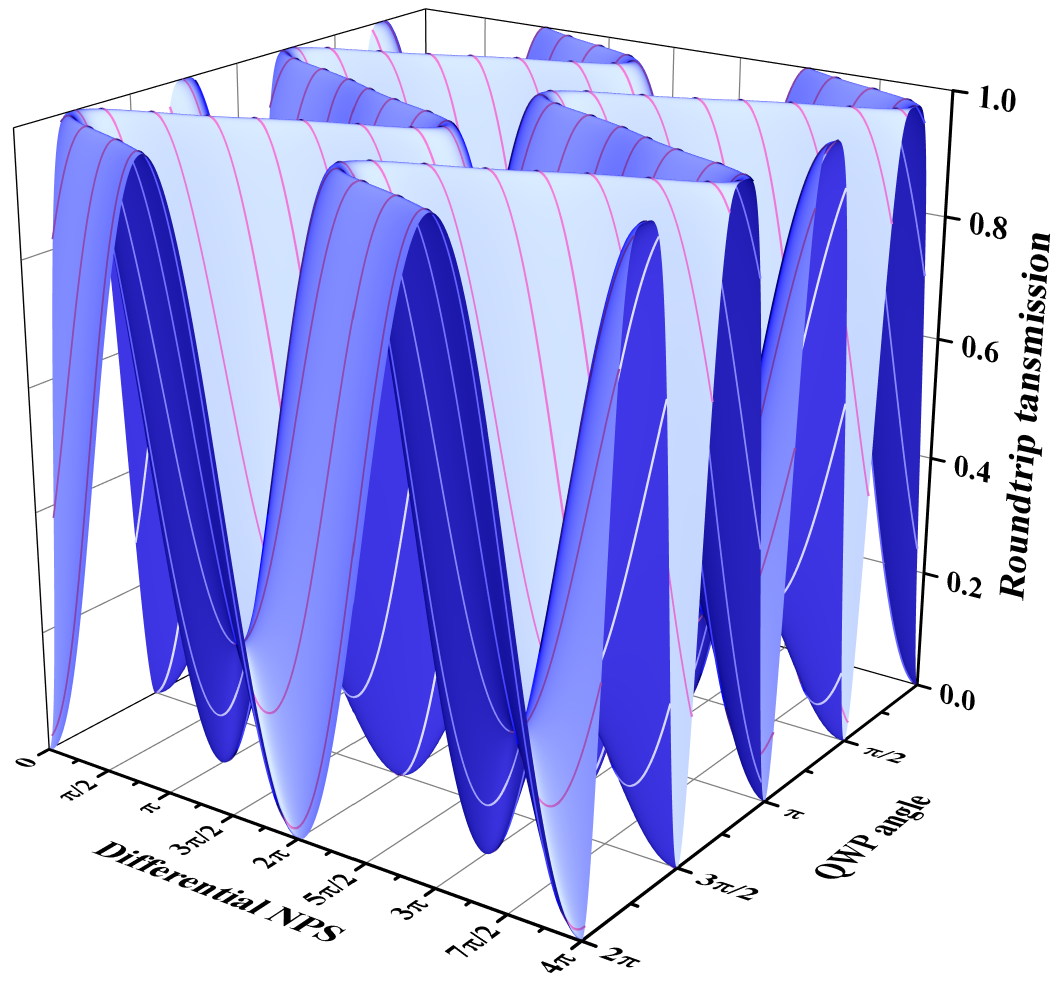}
\caption{The transmission profile for the experimental setup.}
\label{fig_2}
\end{figure}

We choose the transmission of light from the right side of PBS as the starting point for a round trip. Then, the linearly polarized light is reflected by M2, passes through PBS, and is converted into elliptically polarized light after passing through QWP and FR2. It is then split into two beams by PBCC2 and transmitted into different fibers. After the two beams are combined on PBCC1 and output to free space, they are reflected back to PBCC1 with a 90$^\circ$ polarization rotation through FR1, and the two beams enter different optical fibers than before. They are then combined on PBCC2, and the angle between their polarization directions is 90$^\circ$, as shown in Fig.~\ref{fig_1}(b). When two beams pass through the phase shifter and interfere on the PBS, the transmitted light returns to the starting point. The saturable absorption effect of NALM is related to the transmission on PBS after the two beams interfere, which is mainly determined by the differential NPS $\Delta \varphi$ and the QWP angle $\theta$. In order to understand the saturable absorption mechanism of this linear shape laser, the intra-cavity round-trip transmission obtained using the Jones matrix\cite{jones_new_1941,mayer_flexible_2020} is
\begin{align}
T & =\frac{1}{16}\left| 2i\cos \left( 2\theta \right) \left( e^{i\Delta \varphi}-1 \right) \right. 
\notag
\\
& \left. +\left[ 2i\sin \left( 2\theta \right) +\sin \left( 4\theta \right) \right] \left( e^{i\Delta \varphi}+1 \right) \right|^2.
\label{eq1}
\end{align}
The corresponding transmission profile shown in Fig.~\ref{fig_2} has two periods in the QWP angular range of 0--2$\pi$, and the transmission curve is a cosine-like curve for any given QWP angle. As the QWP angle changes, the minimum transmission of the curve also changes, and the width and position of the peak of the curve also undergo periodic shifts. It is important to note that not all curves on the profile can support the mode-locking, and the mode-locking range varies with pumping power and cavity dispersion\cite{li_hysteresis_2022,li_probing_2022}. Therefore, viewed along the QWP-angle axis direction, the transmission curves supporting mode-locking overlap and intertwine. If the Jones matrices of the waveplate (WP), FR1, and M1 are expressed as $M_\mathrm{WP}$, $M_\mathrm{FR}$, and $M_\mathrm{M}$, respectively, then
\begin{align}
M_{\mathrm{WP}}M_{\mathrm{FR}}M_{\mathrm{M}}M_{\mathrm{FR}}M_{\mathrm{WP}} &=M_{\mathrm{FR}}M_{\mathrm{M}}M_{\mathrm{FR}}
\notag
\\
&=\left[ \begin{matrix}
	0&		-1\\
	1&		0\\
\end{matrix} \right] .
\label{eq2}
\end{align}
It is found that the light output from the PBCC1 passes through the WP and Faraday mirror and returns to the PBCC1, with only a change in the polarization direction of the light (counterclockwise rotation by 90$^\circ$). The non-reciprocity introduced by the Faraday rotator cancels out the phase delay of the wave plate in the cavity, resulting in the transmission formula Eq.~(\ref{eq1}) not including the WP-angle term. In the experiment, we take advantage of this fact to tune the number and the time-domain distribution of multi-pulses by rotating WP without affecting the mode-locking. To the best of our knowledge, the method of tuning multi-pulses is reported for the first time in the NALM-based mode-locked laser.

In Fig.~\ref{fig_2}, the modulation depth is maximum when the QWP angle is $n\pi/4$, where $n$ is an integer ranging from 0 to 8. However, in these cases, the mode-locking of the laser cannot be obtained in experiments. This is because the required differential NPS is so large that it cannot reach the first peak of the curve at the QWP angle of 0$^\circ$ (or 2$\pi$) and $\pi$, and the beginning of the curve at the QWP angle of $\pi$/2 and 3$\pi$/2 shows inverse saturable absorption. Therefore, the QWP angles that can achieve mode locking are evenly divided into four intervals segmented by the mentioned angles, where the required differential NPS magnitude for reaching the peak and the modulation depth reach a balance.

\section{Results and Discussion}
\begin{figure}[!t]
	\centering
	\includegraphics[width=0.48\textwidth]{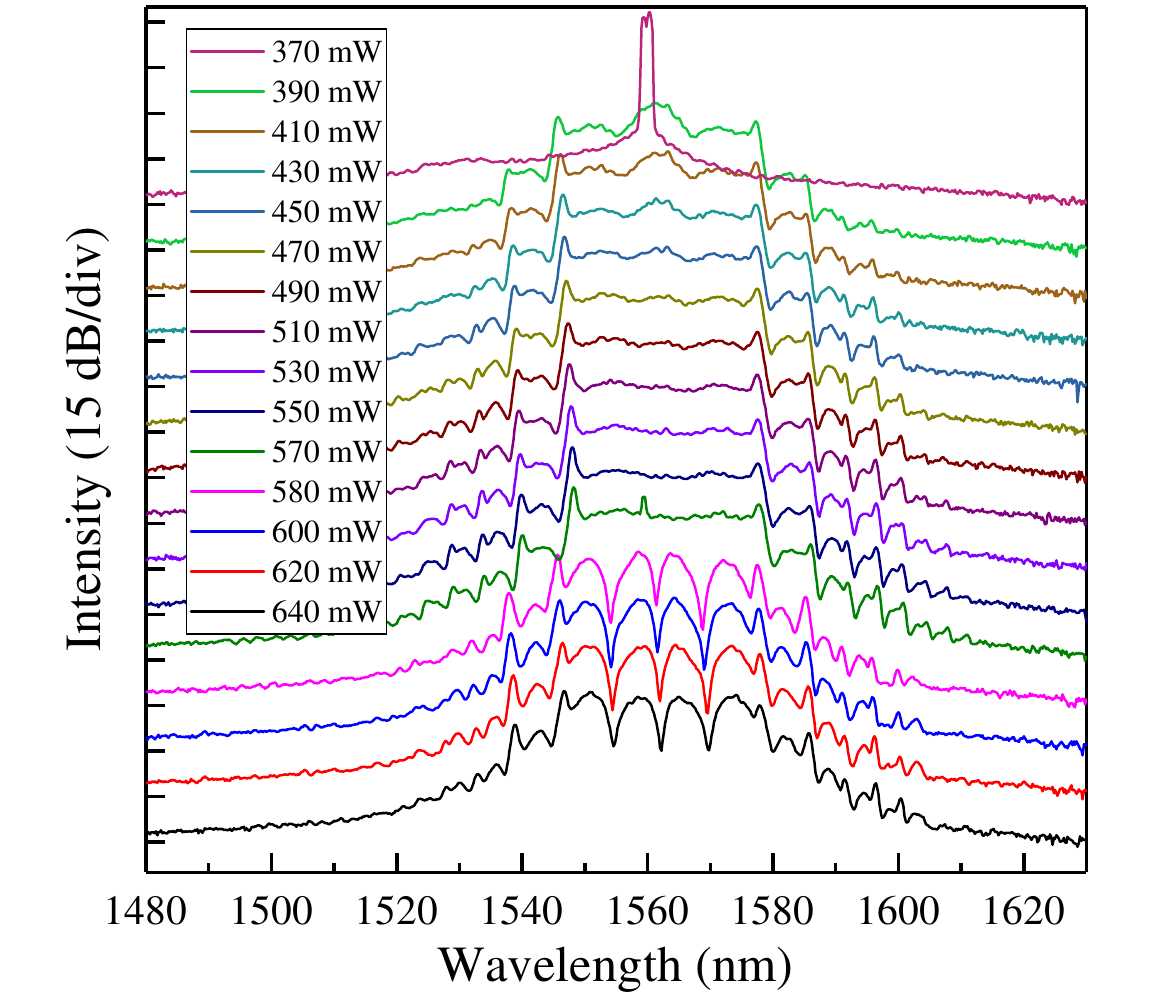}
	\caption{Optical spectra of soliton states and soliton molecule states operating at different pumping powers. The spectral bases are 15 dB apart.}
	\label{fig_3}
\end{figure}

\noindent The mode-locked fiber laser can generate soliton molecules only when the angle of the QWP in the phase shifter is 72$^\circ$ or 108$^\circ$. The spectra of soliton molecules with pumping power from 580 mW to 640 mW are shown in Fig.~\ref{fig_3}, where the peaks between the Kelly sidebands\cite{kelly_characteristic_1992} are consistent with the spectral profiles of the soliton molecules in Ref.~\cite{yun_generation_2012,tang_observation_2001,wang_generation_2016,liu_-demand_2022,gui_soliton_2018} and are the result of the interference between two solitons in the molecule. The possible reason why there are only two angles mentioned here is related to the tangent direction of the surface along the QWP-angle axis in Fig.~\ref{fig_2}. If the angle deviates by about 1$^\circ$, the mode-locked laser will not output soliton molecules and transform into a conventional figure-9 laser\cite{hansel_all_2017}, where high pumping power leads to multi-pulse operation and low pumping power results in single-pulse operation. The pumping threshold ($\sim$ 400 mW) for mode-locking self-starting of this conventional laser is lower than that ($\sim$ 640 mW) of the soliton molecule laser, indicating that the increase in cavity loss is related to the generation of soliton molecules\cite{komarov_multistability_2005}. As the pumping power decreases from 640 mW to 580 mW, the typical soliton-molecule optical spectrum undergoes small changes. The highest peak of the Kelly sidebands on the left shifts towards shorter wavelengths by 0.8 nm, while the highest peak on the right remains relatively unchanged. And the amplitudes of the two peaks close to the spectrum center also slightly increase.

When the pumping power is reduced to a level where the soliton molecule state cannot be sustained, there will be a transition from the soliton molecule state to the soliton state, accompanied by an increase in output power. In the soliton state, a typical Kelly sideband spectrum, as shown in Fig.~\ref{fig_3}, is observed, and the highest peak of the Kelly sidebands on the left also shifts to shorter wavelengths as the pumping power decreases. When transitioning to the soliton state with a pumping power of 570 mW, a peak of continuous light component appears near the center of the spectrum, known as continuous-wave (CW) breakthrough\cite{chouli_soliton_2010,gui_widely_2013,tao_pulse_2023}. This peak disappears as the pumping power is reduced. Interestingly, as the pumping power further decreases, a different peak with increasing intensity appears near the spectral center. At a pumping power of 390 mW, the intensity of this peak surpasses the highest peak of the Kelly sideband by 4.74 dB, and its 3 dB bandwidth (FWHM) also increases to 4.16 nm. Because this peak is broader and well-separated from the previous peak at 570 mW, and it is obtained by reducing the pump power, its main component should be pulsed light rather than CW light. Reducing the pumping power to 370 mW leads to the loss of mode-locking, and the laser output becomes continuous light with a wavelength of 1560 nm. When the QWP angle does not support mode-locking, the wavelength of this CW light will change to 1530 nm, corresponding to the gain spectrum peak of the erbium-doped fiber.

\begin{figure}[!t]
	\centering
	\includegraphics[width=0.42\textwidth]{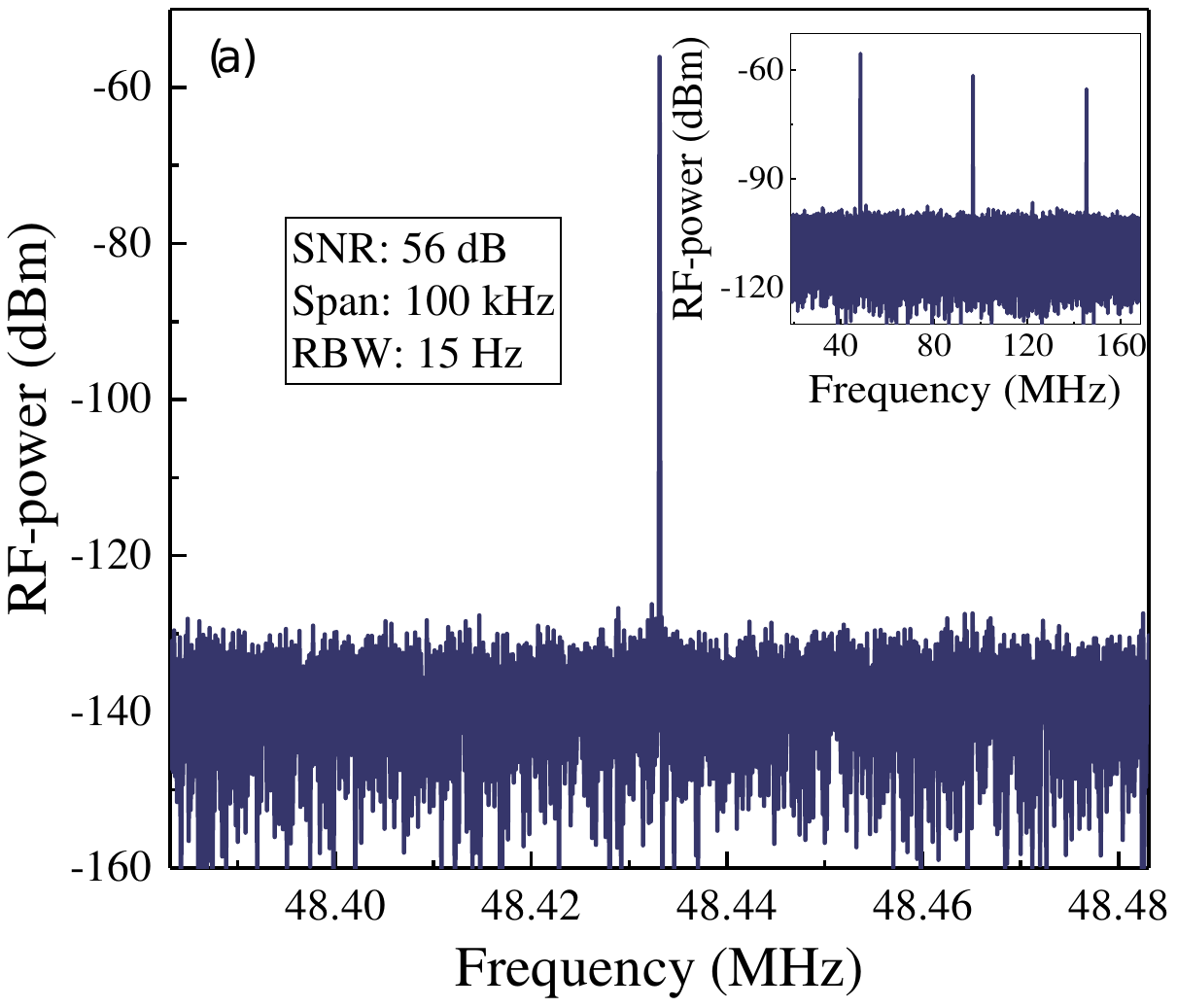}
	\includegraphics[width=0.41\textwidth]{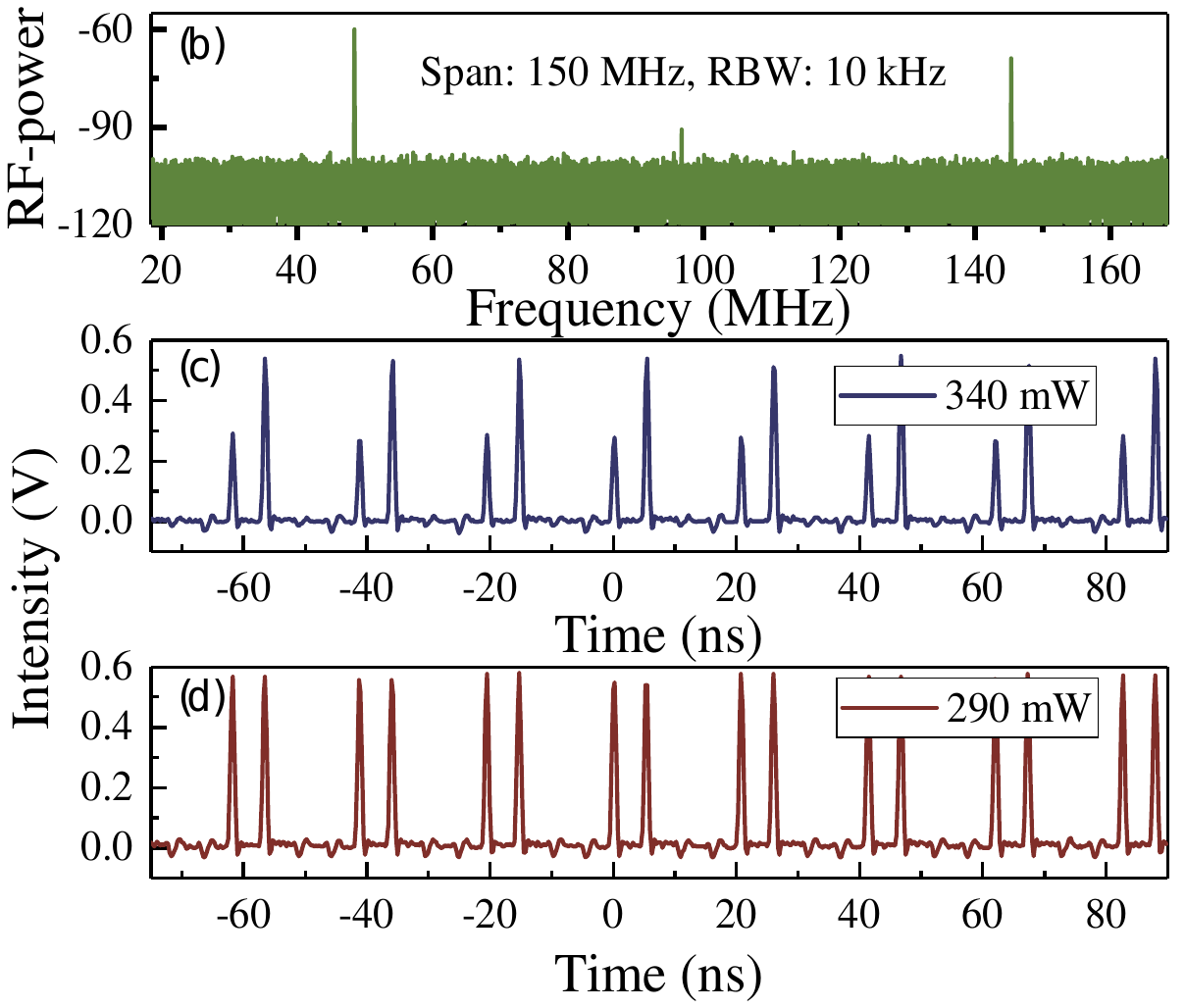}
	\caption{(a) RF spectra of solitons states and soliton molecule states. (b) RF spectra of multi-pulse states, with intensity unit in dBm and SNR of the repetition rate at 59 dB. (c) Oscilloscope trace of multi-pulses at 340 mW. (d) Oscilloscope trace of multi-pulses at 290 mW.}
	\label{fig_4}
\end{figure}

The RF spectra and oscilloscope traces of the soliton state are similar to those of the soliton molecule state. Figure~\ref{fig_4}(a) shows the RF spectra of the soliton molecule state. The oscilloscope trace exhibits a single-pulse pattern. When the pumping power drops to 370 mW, the mode-locking is lost, but it can be achieved again by shaking the PM fiber. 
It is worth noting that, under low pumping power that soliton state cannot be sustained, shaking the PM fiber tends to induce our laser to output multi-pulses rather than single pulses, and the pulse number and oscilloscope traces of the multi-pulses show little variation. Thus, obtaining multi-pulses by shaking the fiber is a repeatable process. 
This phenomenon has not yet been reported in conventional all-PM NALM-based fiber lasers. Even when the same operation is performed in these conventional lasers at low pumping power, only single pulses are obtained. 
In contrast to previous cases to generate multi-pulses by increasing the pumping power\cite{laszczych_mode-locking_2022,guo_single_2022}, the operating pumping powers of the multi-pulse states of our laser are lower than that of soliton states, and the number of pulses can remain constant as the pumping power decreases gradually in a certain range. Figure~\ref{fig_4}(b) shows the RF spectrum of the multi-pulse state at 370 mW. Unlike the RF spectrum shown in Fig.~\ref{fig_4}(a), the second harmonic RF power of the repetition rate is lower. For the multi-pulse states, two types of oscilloscope traces can be observed. One trace consists of a high-intensity pulse and a low-intensity pulse, as shown in Fig.~\ref{fig_4}(c). When the pumping power decreases from 300 mW to 290 mW, a sudden transition from this trace to another trace will occur. The latter trace contains two pulses with similar intensities, as shown in Fig.~\ref{fig_4}(d). Additionally, a slight increase in output power is observed during the transition, indicating that the new stability state experiences low cavity losses.

\begin{figure}[!t]
	\centering
	\includegraphics[width=0.43\textwidth]{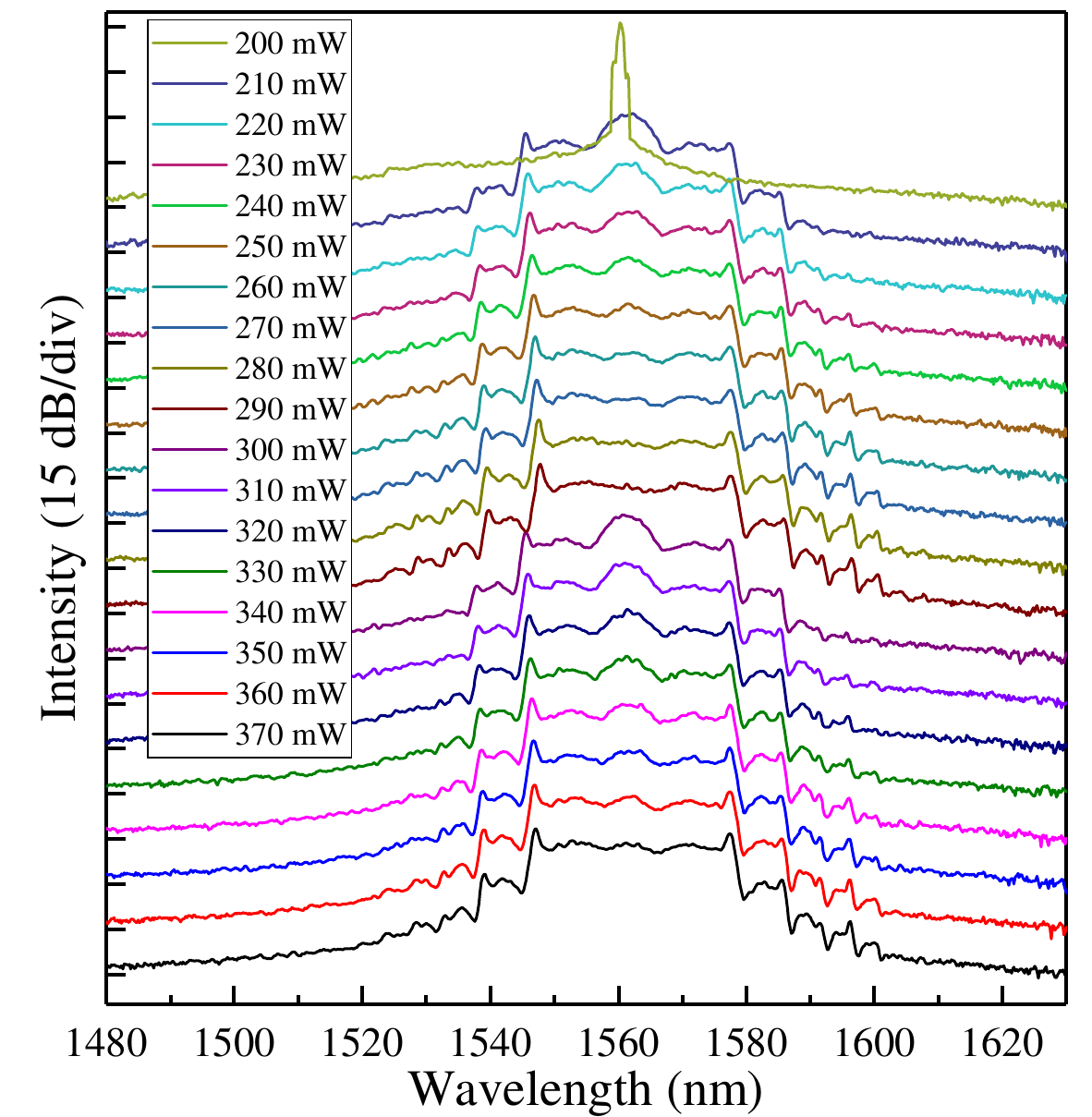}
	\caption{Optical spectra of multi-pulse states operating at different pumping powers. The spectral bases are 15 dB apart.}
	\label{fig_5}
\end{figure}

To analyze the reasons for the sudden trace changes, we measured the multi-pulse spectra at different pumping powers, as shown in Fig.~\ref{fig_5}. The multi-pulse spectra are similar to the single-pulse spectra. Their highest peaks on the left side of the Kelly sidebands also shift to the shorter wavelengths, but the peak shifts back in the sudden trace changes. Because the laser works at a low pumping power, the multi-pulse spectra have little distortion, which differs from the spectra in other works\cite{kobtsev_mode-locked_2018,fan_generation_2018}. In particular, peaks appear in the spectra center, and the peak intensity periodically changes with two cycles as the pumping power decreases. The period of spectral change in the multi-pulse state is shorter than that of the single-pulse state in Fig.~\ref{fig_3}, allowing for the observation of two complete cycles. This is the reason why this periodic phenomenon has not been reported yet. This phenomenon is likely to exist in all mode-locked lasers with different soliton states, not only in single-pulse and multi-pulse states. These two cycles correspond to the oscilloscope traces in Fig.~\ref{fig_4}(c) and Fig.~\ref{fig_4}(d), respectively.

Before interpreting the periodic changes in the spectra in Fig.~\ref{fig_3} and Fig.~\ref{fig_5}, it is necessary to clarify the following experimental facts. Firstly, passively mode-locked lasers have a mechanism to adapt to changes in cavity parameters, enabling them to counteract external disturbances\cite{takeuchi_robust_2023,liao_pulse_2019}. This capability may be associated with the characteristics of passive mode-locking and the broadband nature of mode-locked lasers, where the intensity variations of light at different frequencies in the spectrum can maintain the mode-locking. For example, after changing the pumping power, the pulse spectrum will automatically adjust to adapt to the new parameters. Secondly, reducing pumping power affects the gain and the saturable absorption shown in Fig.~\ref{fig_2}, leading to an increase in cavity losses and a change in pulse width and spectrum. Thirdly, there are multistability and hysteresis phenomena in passively mode-locked fiber lasers\cite{komarov_multistability_2005,tang_mechanism_2005,li_hysteresis_2022}. It indicates that the operation of single stability state can be sustained within a certain range of pumping power. When the pumping disturbance exceeds a threshold, transitions between these stability states can occur. Lastly, the pumping power changes the optical path by influencing nonlinear index coefficient of fiber and then affects the filtering of the cavity to different frequencies. 

Based on these points, we believe that mode-locked lasers have the ability to counteract disturbances, operating at a stability state constructed by nonlinearity, gain, losses, and dispersion. When the pumping power change is introduced, each frequency component and its intensity will be automatically (or passively) adjusted so that the laser still operates in the stability state. As the pumping power gradually decreases, the cost of maintaining the stability state increases. Once disturbances become insurmountable, a transition occurs from the stability state to a new stability state with smaller internal conflicts or an unstable state where the mode-locking is lost. Therefore, the periodic optical spectra in Fig.~\ref{fig_3} and Fig.~\ref{fig_5} is the result of counteracting the disturbances and the transition between stability states. With the pumping power decreasing gradually from 370 mW, a peak appears in the center of the optical spectrum and its intensity increases to maintain the stability state. When the pumping power decreases to 290 mW, the stability state cannot be maintained, and the transition to a new stability state occurs with a slight increase in output power due to reduced losses (or internal conflicts). When the pumping power decreases to 200 mW, the mode-locking cannot be maintained. In addition, when the QWP angle is 72$^\circ$ or 108$^\circ$, rotating the WP cannot adjust the pulse number and oscilloscope trace of the above stable multi-pulses, but it can adjust that of the relatively unstable multi-pulses at high pumping power. Deviating the angle by about 1$^\circ$ also enables adjustment of the pulse number and oscilloscope trace of the multi-pulses in the conventional NALM-based fiber laser.

\section{Conclusion}
\noindent We propose a linear shape, all-PM, NALM-based erbium-doped fiber mode-locked laser. Analysis of the transmittance curve derived from the Jones matrix reveals that the rotation of the intra-cavity wave plate positioned outside the phase shifter does not disrupt the mode-locking. This wave plate allows for fine adjustment of the number and temporal structure of multi-pulses. Soliton molecules can be directly obtained after self-starting mode-locking without any additional adjustment. Subsequently, decreasing the pumping power leads to a transition to single-pulse operation, accompanied by a gradually increasing peak in the spectral center. Below a certain level of pumping power, the mode-locking cannot be sustained. Shaking the fiber at this moment can enable multi-pulse operation. Due to operating at low power, the temporal distribution and the spectral characteristics of multi-pulses remain stable, similar to that of single pulses. In the multi-pulse state, the peak in the spectral center that occurred during single-pulse operation exhibits periodic changes in intensity with the decrease in pumping power. We introduce a multistability model to explain this phenomenon. These findings are helpful for understanding and controlling the stability state transitions of the mode-locked fiber lasers, offering new perspectives for the study of pulse dynamics and pulse state switching.

\bibliography{Reference}

\end{document}